\documentclass[conference]{IEEEtran}
\IEEEoverridecommandlockouts
\usepackage{cite}
\usepackage{threeparttable}
\usepackage{graphicx}
\usepackage{picinpar}
\usepackage[cmex10]{amsmath}
\usepackage{psfrag}
\usepackage{amsopn}
\usepackage[usenames]{color}
\usepackage{enumerate}
\usepackage{subfigure}
\usepackage{epsfig}
\usepackage{amssymb}
\usepackage{amsmath}
\usepackage{bm}
\usepackage{stfloats}

\usepackage{algorithm}
\usepackage{algpseudocode}
\usepackage{amsfonts}
\usepackage{epstopdf}
\usepackage{lipsum}
\def\BibTeX{{\rm B\kern-.05em{\sc i\kern-.025em b}\kern-.08em
    T\kern-.1667em\lower.7ex\hbox{E}\kern-.125emX}}
\begin{document}

\newtheorem{lemma}{Lemma}
\newtheorem{corollary}{Corollary}
\newtheorem{theorem}{Theorem}
\newtheorem{proposition}{Proposition}
\newtheorem{definition}{Definition}

\title{RIS-Aided Offshore Communications with Adaptive Beamforming and Service Time Allocation\\}

\author{\IEEEauthorblockN{Zhengyi Zhou$^1$, Ning Ge$^1$, Wendong Liu$^1$, and Zhaocheng Wang$^{1,2}$}
\IEEEauthorblockA{
$^1$Beijing National Research Center for Information Science and Technology (BNRist),\\
Department of Electronic Engineering, Tsinghua University, Beijing 100084, China \\
$^2$Tsinghua Shenzhen International Graduate School, Shenzhen 518055, China}
}

\maketitle

\begin{abstract}

Reconfigurable intelligent surfaces (RISs),
which can deliberately adjust the phase of incident waves,
have shown enormous potentials to reconfigure the signal propagation for performance enhancement.
In this paper, we investigate the RIS-aided offshore system to provide a cost-effective coverage of high-speed data service. The shipborne RIS is placed offshore to improve the signal quality at the vessels, and the coastal base station is equipped with low-cost reconfigurable reflect-arrays (RRAs),
instead of the conventional costly fully digital antenna arrays (FDAAs), to reduce the hardware cost.
In order to meet the rate requirements of diversified maritime activities,
the effective sum rate (ESR) is studied by jointly optimizing the beamforming scheme and the service time allocated to each vessel.
The optimal allocation scheme is derived, and
an efficient fixed-point based alternating ascent method is developed to obtain a suboptimal solution to the non-convex beamforming problem.
Numerical results show that the ESR is considerably improved with the aid of the RIS,
and the proposed scheme using the hardware-efficient RRAs has only a slight performance loss, compared to its FDAA-based counterpart.

\end{abstract}


\section{Introduction}
Due to the rapid development of blue economy, the demand for high-speed data services of diversified maritime activities has been continuously increasing over the past years \cite{Mari1}.
The conventional MF/HF/VHF band systems offer a wide coverage, but cannot support high-speed data transmission with limited bandwidth, while the satellite systems are cost-ineffective and also impose a relatively large latency.
Leveraging higher frequency bands, offshore communications have drawn much attention for its potential to provide the vessels with broadband service and wide coverage (up to 100 km)  \cite{Mari2}.

At present, the state-of-the-art solutions tried to introduce the technology used in terrestrial communications such as Wi-Fi, WiMax, and LTE, into offshore systems with the aid of coastal base stations (BSs)
\cite{Mari2, Mari3, Mari4, Mari5}.
With adequate transmission power, the downlink data rate could be increased to Mbps level at a distance of dozens of kilometers between the vessel and the coast \cite{Mari1,Mari3,Mari4}.
However, due to the considerable path loss and the geographic limitation on the sea,
it is still costly and power-consuming to provide a favorable signal quality at the receiver for high-speed data service.
How to extend the coverage area and minimize the networking expense is an ongoing and challenging problem.

Recently, a new concept of reconfigurable intelligent surfaces (RISs) has emerged.
A RIS is a planar array consisting of many low-cost elements that only reflect the incident waves with an adjustable phase shift.
By properly modifying the phase shift,
the signal propagation can be reconfigured in a deliberate manner
without extra operations for baseband/radio frequency (RF) processing or retransmission \cite{RIS1,RIS2,RIS3,RIS4, RIS5}.
It is expected that, with the aid of RISs in offshore systems, the signal quality at the vessels can be substantially improved with very little additional power consumption,
and hence the coverage area can be much extended at a relatively low cost.

The applications of RISs in terrestrial systems have been investigated in many works.
Most studies considered a downlink scenario, where the RIS is placed between the BS and the users to establish a BS-RIS-user reflected link in addition to the conventional BS-user direct link for performance enhancement.
In \cite{RIS2}, the ergodic spectral efficiency of the RIS-assisted reflected link was evaluated,
and an optimal phase shift design was developed based on statistical channel state information (CSI).
The authors of \cite{RIS3} also focused on the reflected link, and optimized the transmit power allocation at the BS.
The joint beamforming scheme of both the direct and reflected links
was derived for different objectives, including minimizing the transmit power \cite{RIS4}, and maximizing the weighted sum rate \cite{RIS5}.
Current works have all considered digital beamforming with fully digital antenna arrays (FDAAs) at the BS.
However, the implementation of FDAAs requires one dedicated RF chain for each antenna element,
which causes extremely high hardware cost and power consumption for large-scale antenna arrays \cite{FDAA}.
Since beamforming with large arrays is anticipated to support long-distance maritime communication,
it is not practical to use conventional FDAAs in offshore systems.
A more cost-effective scheme is to apply hybrid or analog beamforming with only a limited number of RF chains,
while due to the non-convex constraints imposed by both the RF beamformer and the RIS elements,
the joint beamforming design of the direct and reflected links becomes even more difficult.
Also, because of the large path loss deviation among different vessels in the wide sea area,
serving multiple vessels simultaneously requires extremely high linearity for power amplifiers.
Hence, it is practical to let the BS serve a single vessel at one time with full transmission power to surmount the large path loss,
and adaptively allocate its service time to different vessels for their individual rate requirements.
The joint optimization over the service time allocation and the beamforming design
has not yet been considered for RIS-aided wireless systems, which thus motivates this work.

In this paper, we investigate the RIS-aided offshore maritime system for a cost-effective coverage of broadband service.
The shipborne RIS is placed in costal waters to improve the signal quality at the vessels,
and is controlled by the coastal BS equipped with large-scale antenna arrays.
The BS adaptively allocates its service time to the vessels,
and serves a single vessel at one time.
Instead of the conventional costly FDAAs,
low-cost reconfigurable reflect-arrays (RRAs) are employed to reduce the hardware cost.
In order to meet the diversified requirements of maritime activities,
we define the effective sum rate (ESR)
as the sum of each vessel's rate weighted by
the service time allocated by the BS, and
maximize the ESR under the vessels' individual rate constraints.
The allocation scheme is jointly optimized together with the beamforming design,
which differs from the weighted sum rate optimization with predetermined weights in \cite{RIS5}.
The optimal allocation design is derived, and an efficient fixed-point based method is developed for the joint beamforming problem.
The performance of the proposed RIS-assisted system is evaluated through extensive numerical experiments.
Simulation results show that the ESR is considerably improved with the aid of the RIS,
and the RRA-based scheme with a much lower cost, has only a slight performance loss, compared to its FDAA-based counterpart.

\section{System Model}

\subsection{Signal Model}

\begin{figure}[t!]
\begin{center}
\includegraphics[width=1\linewidth]{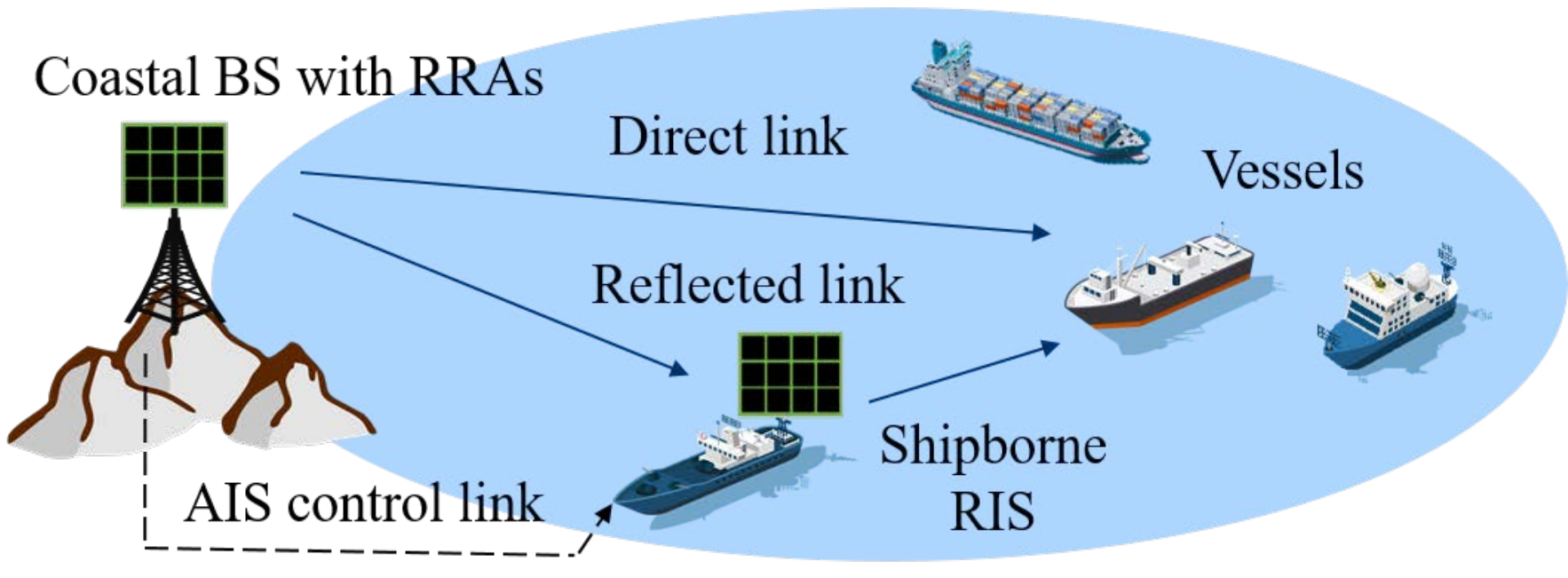}
\end{center}
\vspace*{-6mm}
\caption{RIS-aided offshore communication systems.}
\label{Fig1}
\vspace*{-1mm}
\end{figure}

 As shown in Fig.~\ref{Fig1},
 we consider a single-cell scenario in the RIS-aided offshore system,
 where the remote RIS is shipborne and placed in the coastal waters to assist with the signal transmission from the multi-antenna BS on the land to $K$ single-antenna vessels.
 The RIS consists of $M$ reflecting elements, and the BS is equipped with RRAs of $N$ elements.
 A controller is mounted at the RIS,
 and a wireless control link based on automatic identification system (AIS) is leveraged by the BS to coordinate the behavior of the RIS.\footnote{AIS is a conventional VHF-band maritime system that can provide a 120 km communication coverage with a maximum data rate of 9.6 kbps \cite{Mari4}.
 AIS data contain instantaneous information of ships such as position and velocity, and are generally used for the safety of ships \cite{Mari1,Mari2}.}
 To improve the signal quality at the vessel, the BS allocates an exclusive service time for every vessel, and provides a single independent data stream to serve one of them at one time with full transmission power.

 For the $k$th vessel, the allocated service time is denoted by $t_k(t_k \geq 0)$, and the transmit symbol is expressed as $s_k\in
 \mathbb{C}$ with $\textsf{E}\big\{s_ks_k^{\rm H}\big\}=1$.
 Two diagonal matrices, $\bm{\Lambda}_k=\text{diag}\left\{e^{\textsf{j}v_{1,k}}, e^{\textsf{j}v_{2,k}},
 \cdots ,e^{\textsf{j}v_{N,k}}\right\}$, and
 $\bm{\Sigma}_k=\text{diag}\left\{e^{\textsf{j}u_{1,k}}, e^{\textsf{j}u_{2,k}},
 \cdots ,e^{\textsf{j}u_{M,k}}\right\}$,
 are defined to denote the adjustable phase shifts implemented by the RRA and RIS elements for the $k$th vesesel, respectively,
 where $v_{n,k},  u_{m,k} \in [0,2\pi)$, and $\textsf{j}$ is the imaginary unit.

 The superposed signals of the direct and reflected links received at the $k$th vessel can be represented by \cite{RIS2,RIS3,RIS4,RIS5}
\begin{equation}\label{Eq:Rx_signal}
 y_k = \sqrt{\rho}\Big(\sqrt{\delta_k^{(\rm d)}} \bm{\bar{h}}_k^{\rm H} + \sqrt{\delta_k^{(\rm r)}} \bm{\bar{g}}_k^{\rm H} \bm{\Sigma}_k \bm{F}\Big)\bm{\Lambda}_k \bm{r} s_k + \xi_k,
\end{equation}
 where $\rho$ is the transmission power, $\xi_k$ is the noise following a complex symmetric Gaussian
 distribution $\xi_k \sim \mathcal{CN}(0, \sigma^2)$,
 $\bm{r} \in \mathbb{C}^{N\times 1}$ is a parameter of the RRAs,
 and $\delta_k^{(\rm d)} (\delta_k^{(\rm r)})$ is the path loss of the direct(reflected) link,
 while $\bm{\bar{h}}\in \mathbb{C}^{N\times 1}$, $\bm{F}\in \mathbb{C}^{M\times N}$,
 and $\bm{\bar{g}}\in \mathbb{C}^{M\times 1}$
 denote the channels from the BS to the vessel, from the BS to the RIS, and from the RIS to the vessel, respectively.

\begin{figure}[t!]
\begin{center}
\includegraphics[width=.9\linewidth]{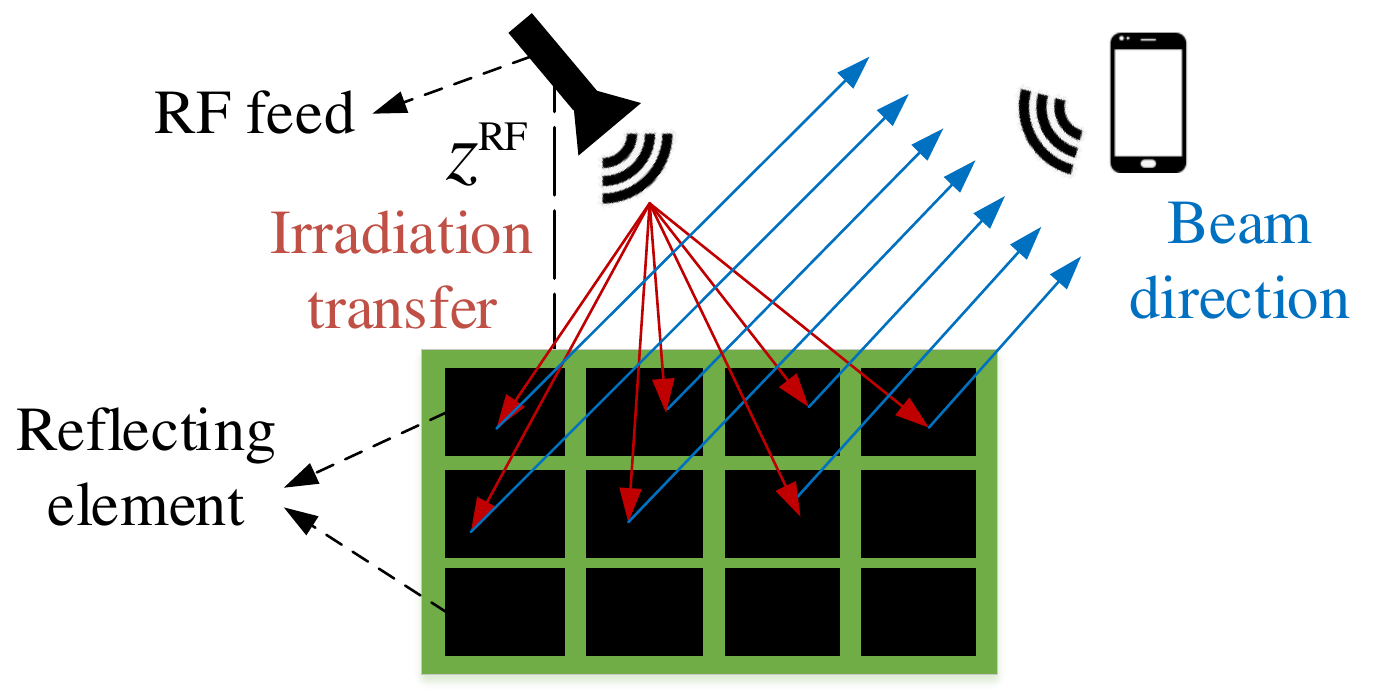}
\end{center}
\vspace*{-6mm}
\caption{The structure of RRAs with one RF feed.}
\label{Fig2}
\vspace*{-1mm}
\end{figure}

 The structure of the RRAs is illustrated in Fig.~\ref{Fig2}.
 The transmitted signal is transferred from the RF feed to the array surface via irradiation,
 wherein an amplitude variation of $|r_i|$ and a phase delay of $\angle{r_i}$ are introduced before the signal reaches the $i$th reflecting elements.
 According to the feed horn radiation pattern model, the values of $|r_i|$ and $\angle{r_i}$ are determined by
\begin{align}\label{eq4}
 \angle{r_i} =& 2\pi S_i/\lambda , \quad |r_i| = \gamma \cos^q(\omega_i) ,
\end{align}
 where $S_i$ is the distance between the RF feed and the $i$th element,
 $q$ is a system parameter, $\gamma$ is a normalizing scalar, and $\omega_i=\arccos\big(z^{\rm RF}/S_i\big)$
 is the elevation angle between the RF feed and the $i$th element with $z^{\rm RF}$ denoting the height of the RF feed \cite{RA1,RRA2}.
 The irradiation transfer process is specified by the complex vector $\bm{r}\in\mathbb{C}^{N\times 1}$,
 with entries of $r_i=|r_i| e^{\textsf{j}\angle{r_i}}$.
 The transfer vector is normalized to $\|\bm{r}\|_2=1$,
 and is considered to be constant for a well fabricated array.

\subsection{Channel Model}
The typical two-ray propagation model is utilized to calculate the path loss on the sea.
The loss function is given by
 \begin{equation}\label{Eq:pathloss}
 L(l) = (\frac{\lambda}{4\pi l})^2\times \sin^2(\frac{2\pi h_th_r}{\lambda l}),
 \end{equation}
 where $l$ is the distance between the transmitter and the receiver, $\lambda$ is the transmission wavelength, and $h_t(h_r)$ is the height of the transmitting(receiving) antenna \cite{Mari2,Mari3}.
 Given the propagation distances of the direct and reflect links,
 the parameters defined in (\ref{Eq:Rx_signal}) are determined by $\delta_k^{(\rm d)}=L(l^{(\rm d)})$ and $\delta_k^{(\rm r)}=L(l^{(\rm r)})$, where $l^{(\rm d)}$ is the distance from the BS to the vessel, and $l^{(\rm r)}$ is the sum of the distance from the BS to the RIS and that from the RIS to the vessel \cite{RIS_PathLoss}.
 For small-scale fading, the Rician model is used for all the involved channels.
 Hence, the channel $\bm{F}$ is expressed as \cite{RIS2,RIS3}
 \begin{equation}\label{Eq:Channel}
 \bm{F} = \sqrt{\frac{\alpha}{1+\alpha}}\bm{F}^{(\rm LOS)}  + \sqrt{\frac{1}{1+\alpha}} \bm{F}^{(\rm NLOS)},
 \end{equation}
 where $\alpha$ is the Rician factor, $\bm{F}^{(\rm LOS)}$ is the line-of-sight (LOS) component,
 and $\bm{F}^{(\rm NLOS)}$ is the non-LOS (NLOS) component.
 The entries of $\bm{F}^{(\rm NLOS)}$ are i.i.d., and follow the complex symmetric Gaussian distribution with zero mean and unit variance,
 and $\bm{F}^{(\rm LOS)}$ is given by
 \begin{align}\label{Eq:channel} 
 \bm{F}^{(\rm LOS)} = \bm{a}_\text{r}\big(\phi^{(\rm r)},\varphi^{(\rm r)}\big)
  \bm{a}_\text{t}^{\rm H}\big(\phi^{(\rm t)},\varphi^{(\rm t)}\big) ,
\end{align}
 where $\phi^{(\rm r)}$ ($\varphi^{(\rm r)}$) and
 $\phi^{(\rm t)}$ ($\varphi^{(\rm t)}$) are the azimuth (elevation) angles of
 arrival and departure, and $\bm{a}_{\rm r}
 (\phi,\varphi )$ ($\bm{a}_{\rm t}(\phi,\varphi )$) are the transmit (receive) antenna array
 response vectors.
 For a uniform planar array (UPA) with size of $N_\text{h} \times N_\text{v}$, the array
 response vector $\bm{a}_{\rm UPA} (\phi,\varphi )$ can be expressed as
\begin{align}\label{Eq:steering} 
 \bm{a}_{\text{UPA}}(\phi,\varphi ) =& \frac{1}{\sqrt{N_\text{h} N_\text{v}}} \Big[ 1 \cdots
  e^{\textsf{j}\frac{2\pi}{\lambda} d (m\sin\phi\sin\varphi + n\cos\varphi)} \cdots \nonumber \\
 &  e^{\textsf{j}\frac{2\pi}{\lambda} d ((N_\text{h}-1)\sin\phi\sin\varphi + (N_\text{v}-1)\cos\varphi)}\Big]^{\rm T} ,
\end{align}
 where $d$ is the spacing between adjacent array elements, $0\le m\le (N_\text{h}-1)$, and $0\le n\le (N_\text{v}-1)$.
 The channels $\bm{\bar{h}}$ and $\bm{\bar{g}}$ are derived in the same way, and the expressions are omitted for simplicity.

Channel estimation is a challenging issue for the practical applications of RISs.
With no RF chains mounted,
it is infeasible to directly estimate the channels $\bm{F}$ and $\bm{\bar{g}}$ at the remote RIS.
However, due to the LoS-dominant propagation feature in costal waters,
large-scale CSI, such as path loss, and angle-of-arrival/departure (AoA/AoD),
can be provided by the AIS link to assist with the beamforming design.
The problems of channel estimation and beamforming based on imperfect CSI are discussed in future works.
In this paper, complete CSI is assumed, which is the same as \cite{RIS3,RIS4,RIS5}.

\section{Proposed RIS-Aided Near Coast Maritime Communication Systems}
Given the service time of $t_k$, the communication rate for the $k$th vessel is obtained as
\begin{equation}\label{Eq:R}
 R_k = \log_2\Big(1 + \frac{\rho}{\sigma^2}  \big| (\bm{h}_k^{\rm H} + \bm{g}_k^{\rm H} \bm{\Sigma}_k \bm{F})\bm{\Lambda}_k \bm{r} \big|^2 \Big),
\end{equation}
where $\bm{h}_k = \delta_k^{(\rm d)} \bm{\bar{h}}_k$, and $\bm{g}_k=\delta_k^{(\rm r)}  \bm{\bar{g}}_k$.
Then the rate constraint is expressed by
\begin{equation}
 t_kR_k \geq \beta_k,
\end{equation}
where $\beta_k$ is the individual rate requirement of the $k$th vessel, and the service time is normalized such that $\sum_{k=1}^K t_k = 1$.

The constrained maximization of the ESR can be posed as the following problem,
\begin{equation}\label{Eq:OP1} 
\begin{aligned}
 \!\!\!   (\text{P1}): \underset{ t_k, \bm{\Lambda}_k, \bm{\Sigma}_k}  \max \  &
   \text{ESR}= \underset{k=1} {\overset{K} \sum } \ t_kR_k, \\
\text{s.t.} \quad
 & \!\!\! \bm{\Lambda}_k \in \mathcal{D}^N \triangleq \left\{\bm{\Lambda}=\text{diag}
  \left\{e^{\textsf{j}v_1},\cdots ,e^{\textsf{j}v_N}\right\}\right\}, \ \forall k, \\
 & \!\!\!  \bm{\Sigma}_k \in \mathcal{D}^M \triangleq \left\{\bm{\Sigma}=\text{diag}
  \left\{e^{\textsf{j}u_1},\cdots ,e^{\textsf{j}u_M}\right\}\right\}, \ \forall k, \\
 & \!\!\! t_k R_k \geq \beta_k, \ \forall k,\\
 & \!\!\!  \underset{k=1} {\overset{K} \sum } \ t_k = 1, \quad t_k \geq 0,  \ \forall k.
 \end{aligned}
\end{equation}
It is noted that the existence of the solution to (P1) is not guaranteed.
Intuitively, given the wireless channels and the transmission power of the BS,
the achieved communication rate $R_k$ is always bounded.
For excessively large values of $\beta_k$,
the constraints of $t_kR_k \geq \beta_k$ cannot be satisfied whatever service time allocation  or beamforming design is leveraged.
Without loss of generality, we assume that the values of $\beta_k$ are proper so that the feasible domain of (P1) is not empty.
In the sequel of this section, we derive a sufficient and necessary condition for the feasibility of (P1).

The problem is non-convex due to the constraints that $\bm{\Lambda}_k$ and $\bm{\Sigma}_k$ have to be diagonal and unitary.
In the next subsections, we first derive the optimal values for $t_k$ by analyzing the complementary slackness conditions,
and then develop an effective fixed-point based alternating ascent method to acquire the suboptimal solutions for $\bm{\Lambda_k}$ and $\bm{\Sigma}_k$.

\subsection{Optimal Allocation of the Service Time }\label{AA}

By change of variables, (P1) can be reformulated as follows
\begin{equation}\label{Eq:OP2} 
\begin{aligned}
 \!\!\!  (\text{P2}): \underset{ t_k,\bm{\mu}_k, \bm{b}_k}  { \max} &\
   \underset{k=1} {\overset{K} \sum } \ t_k\log_2\Big(1 + \frac{\rho}{\sigma^2}  \big|(\bm{A}_k \bm{\mu}_k + \bm{c}_k)^{\rm H}\bm{b}_k \big|^2 \Big), \\
\text{s.t.} \quad & \!\!\! t_k \log_2\Big(1 + \frac{\rho}{\sigma^2} \big|(\bm{A}_k \bm{\mu}_k + \bm{c}_k)^{\rm H}\bm{b}_k \big|^2 \Big) \geq \beta_k, \ \forall k,\\
 & \!\!\!  |\mu_{i,k}| = 1, \quad |b_{i,k}| = 1, \ \forall i,k, \\
 & \!\!\!  \underset{k=1} {\overset{K} \sum } \ t_k = 1, \quad t_k \geq 0,  \ \forall k.
 \end{aligned}
\end{equation}
where
$\bm{R} = \text{diag}(\bm{r})$,
$\bm{G}_k = \text{diag}(\bm{g}_k)$,
$\bm{c}_k=\bm{R}^{\rm H}\bm{h}_k$,
$\bm{A}_k = \bm{R}^{\rm H} \bm{F}^{\rm H} \bm{G}_k$,
$\bm{b}_k = [b_{1,k} \cdots b_{N,k}]^{\rm T} = \text{diag}(\bm{\Lambda}_k) $,
and $\bm{\mu}_k = [\mu_{1,k} \cdots \mu_{M,k}]^{\rm T}= \text{diag}(\bm{\Sigma}_k^{\rm H}) $.

 The complementary slackness condition with respect to the rate constraints in (\ref{Eq:OP2}) is given by
 $\lambda_k(\hat{t}_k\hat{R}_k-\beta_k)=0$,
 where $\lambda_k$ is the lagrangian multiplier,
 $\hat{t}_k$ is the optimal solution of $t_k$,
 and $\hat{R}_k$ is the rate achieved by the optimal $\bm{\mu}_k$ and $\bm{b}_k$.
 Thus, the sufficient condition for the optimal allocation of the service time can be derived by the following proposition.
\begin{proposition}\label{Prop:1}
 If $(\hat{t}_k, \hat{\bm{\mu}}_k, \hat{\bm{b}}_k)$ solves (P2), then
\begin{equation}\label{Eq:t}
\hat{t}_k =
\left\{\begin{aligned}
  \beta_k / \hat{R}_k, \quad \quad &\text{if} \quad k \not= k^*,\\
 1- \sum_{k \not=k^*}\beta_k / \hat{R}_k,  \ \ & \text{if} \quad k = k^*,
\end{aligned}\right.
\end{equation}
where $\hat{R}_k = \log_2\Big(1 + \frac{\rho}{\sigma^2} \big|(\bm{A}_k \bm{\hat{\mu}}_k + \bm{c}_k)^{\rm H}\bm{\hat{b}}_k \big|^2 \Big)$,
and $k^* = \arg \underset{k}\max \ |(\bm{A}_k \bm{\hat{\mu}}_k + \bm{c}_k)^{\rm H}\bm{\hat{b}}_k| $.
\end{proposition}

\begin{IEEEproof}
 For the case of $k\not=k^*$, without loss of generality, assume that $\exists k_0(k_0 \not=k^*)$, s.t. $\Delta t = \hat{t}_{k_0} - \beta_k/\hat{R}_k > 0$. Further define $\bar{t}_k$ for $k = 1, \dots, K$ as follows
\begin{equation}
\bar{t}_k =
\left\{\begin{aligned}
  \bar{t}_k = \hat{t}_k,  \quad &\text{if} \quad k \not= k_0 \ \text{or}\ k^*,\\
 \hat{t}_{k_0} - \Delta t,  \ \ & \text{if} \quad k = k_0,\\
 \hat{t}_{k^*} + \Delta t,  \ \ & \text{if} \quad k = k^*.
\end{aligned}\right.
\end{equation}
By comparing the ESR values obtained by $\hat{t}_k$ and $\bar{t}_k$, we have
$\sum_k \bar{t}_k\hat{R}_k - \sum_k \hat{t}_k\hat{R}_k = \Delta t (\hat{R}_{k^*}-\hat{R}_{k_0}) > 0$,
which violates the assumption that $\hat{t}_k$ is the solution to (P2).
Thus, we have $\hat{t}_k = \beta_k / \hat{R}_k$ for $k \not= k^*$,
and $\hat{t}_{k^*} = 1-  \sum_{k \not=k^*}\beta_k / \hat{R}_k$ by the constraint $\sum_k t_k = 1$.
\end{IEEEproof}
A corollary can be directly derived from Proposition \ref{Prop:1}:
\begin{corollary}\label{Coro:1}
 Given $(\bm{\mu}_k, \bm{b}_k)$, the maximum objective value of (P2) is expressed by
\begin{equation}\label{Eq:Coro1}
\begin{aligned}
 \widehat{\text{ESR}}(& \bm{\mu}_k,  \bm{b}_k) = \bigg[1- \sum_{k \not=k^*}  \frac{\beta_k}{\log_2(1 + \frac{\rho}{\sigma^2} |(\bm{A}_k \bm{\mu}_k + \bm{c}_k)^{\rm H}\bm{b}_k |^2 )}\bigg] \\
 & \times \log_2\Big(1 + \frac{\rho}{\sigma^2} \big|(\bm{A}_{k^*} \bm{\mu}_{k^*} + \bm{c}_{k^*})^{\rm H}\bm{b}_{k^*} \big|^2 \Big) + \sum_{k \not=k^*} \beta_k ,
\end{aligned}
\end{equation}
where $k^* = \arg \underset{k}\max \big|(\bm{A}_k \bm{\mu}_k + \bm{c}_k)^{\rm H}\bm{b}_k\big| $.
\end{corollary}
\begin{IEEEproof}
The proof is completed by substituting $t_k$ with $\hat{t}_k$ specified in (\ref{Eq:t}).
\end{IEEEproof}
\subsection{Joint Beamforming of the RIS and the RRAs}
It is observed from (\ref{Eq:Coro1}) that
the optimal value of the ESR is monotonically increasing with $|(\bm{A}_k \bm{\mu}_k + \bm{c}_k)^{\rm H}\bm{b}_k |$.
Therefore, another proposition for the optimal $\bm{b}_k$ is derived as follows.
\begin{proposition}\label{Prop:2}
 If $(\hat{t}_k, \hat{\bm{\mu}}_k, \hat{\bm{b}}_k)$ solves (P2), then
\begin{equation}\label{Eq:b}
\hat{\bm{b}}_k = e^{\textsf{j}\big[\angle(\bm{A}_k \bm{\hat{\mu}}_k + \bm{c}_k)+\theta \bm{1}\big]}
\end{equation}
where $\angle(\cdot)$ is the entry-wise argument operation, and $\theta \in \mathbb{R}$ is any constant number.
\end{proposition}
\begin{IEEEproof}
Denoting $\bm{\hat{w}}_k = \bm{A}_k \bm{\hat{\mu}}_k + \bm{c}_k$, by the triangular inequality and the constraint of $|\hat{b}_{i,k}|=1$, we have
\begin{equation}
|\bm{\hat{w}}_k^{\rm H}\bm{\hat{b}}_k| \leq \sum_k |\hat{w}_{i,k}^*\hat{b}_{i,k} | \leq \|\bm{\hat{w}}_k\|_1,
\end{equation}
where $\|\cdot\|_1$ is the notation for $l_1$-norm.
The equality holds when $\hat{b}_{i,k} = e^{\textsf{j}(\angle\hat{w}_{i,k}+\theta)}$.
Any other $\bm{b}_k$ different from (\ref{Eq:b}) cannot give a larger ESR value, which thus completes the proof.
\end{IEEEproof}

Based on Proposition \ref{Prop:2}, a corollary is obtained as:
\begin{corollary}\label{Coro:2}
 Given $ \bm{\mu}_k$, the maximum objective value of (P2) is expressed by
\begin{equation}\label{Eq:Coro2}
\begin{aligned}
 \widetilde{\text{ESR}}(\bm{\mu}_k) &= \Bigg[1- \sum_{k \not=k^*}  \frac{\beta_k}{\log_2\big(1 + \frac{\rho}{\sigma^2} \|(\bm{A}_k \bm{\mu}_k + \bm{c}_k)\|_1^2 \big)}\Bigg] \\
 & \times \log_2\Big(1 + \frac{\rho}{\sigma^2} \|(\bm{A}_{k^*} \bm{\mu}_{k^*} + \bm{c}_{k^*})\|_1^2 \Big) + \sum_{k \not=k^*} \beta_k,
\end{aligned}
\end{equation}
where $k^* = \arg \underset{k}\max \ \|\bm{A}_k \bm{\mu}_k + \bm{c}_k\|_1 $.
\end{corollary}
\begin{IEEEproof}
The proof is completed by substituting $(t_k, \bm{b}_k)$ with $(\hat{t}_k,\hat{\bm{b}}_k)$ specified in (\ref{Eq:t}) and (\ref{Eq:b}).
\end{IEEEproof}

From Corollary \ref{Coro:2}, the maximized ESR is monotonically increasing with $\|\bm{A}_k \bm{\mu}_k + \bm{c}_k\|_1$.
Therefore, the optimal $\bm{\mu}_k$ for (P2) can be obtained by solving the following problem
\begin{equation}\label{Eq:OP3} 
\begin{aligned}
 \!\!\!  (\text{P3}): \quad  \underset{ \bm{\mu}_k}  \max \  &
   \|\bm{A}_k \bm{\mu}_k + \bm{c}_k \|_1^2, \\
\text{s.t.} \quad
 & \!\!\!  |\mu_{i,k}| = 1, \ \forall i.
 \end{aligned}
\end{equation}
Also, with the solution of (P3), the condition for the feasibility of (P1) and (P2) can be given by the following corollary.
\begin{corollary}\label{Coro:3}
 Given $ \bm{\hat{\mu}}_k$ that solves (P3), (P2) is feasible if and only if
\begin{equation}\label{Eq:Coro13}
 \sum_{k =1}^K  \frac{\beta_k}{\log_2\big(1 + \frac{\rho}{\sigma^2} \|\bm{A}_k \bm{\hat{\mu}}_k + \bm{c}_k\|_1^2 \big)} \leq 1.
\end{equation}
\end{corollary}
\begin{IEEEproof}
Denote the feasible domain of (P2) as $\mathcal{F}^\text{(P2)}$.
On one hand, if (\ref{Eq:Coro13}) holds, then $(\hat{t}_k, \hat{\bm{\mu}}_k, \hat{\bm{b}}_k)\in \mathcal{F}^\text{(P2)}$,
where $\hat{t}_k$ and $\hat{\bm{b}}_k$ are given by (\ref{Eq:t}) and (\ref{Eq:b}). Hence, (P2) has a non-empty feasible domain.

On the other hand, if (P2) is feasible, i.e. $\exists (t_k, \bm{\mu}_k, \bm{b}_k)\in \mathcal{F}^\text{(P2)}$,
s.t. $t_k \geq \beta_k / R_k $, then we have
\begin{equation}
 \sum_{k =1}^K  \frac{\beta_k}{\log_2\big(1 + \frac{\rho}{\sigma^2} \big|(\bm{A}_k \bm{\mu}_k + \bm{c}_k)^{\rm H}\bm{b}_k \big|^2 \big)} \leq 1.
\end{equation}
By Proposition \ref{Prop:2} and the fact that $ \bm{\hat{\mu}}_k$ solves (P3),
we have $|(\bm{A}_k \bm{\mu}_k + \bm{c}_k)^{\rm H}\bm{b}_k | \leq \|\bm{A}_k \bm{\hat{\mu}}_k + \bm{c}_k)\|_1$.
Thus, (\ref{Eq:Coro13}) holds, and the proof is completed.

\end{IEEEproof}

Both the objective and the constraints of (P3) are non-convex.
Therefore, we reformulate the problem as
\begin{equation}\label{Eq:OP4} 
\begin{aligned}
 \!\!\!  (\text{P4}): \quad  \underset{ \bm{\mu}, \bm{b}, \bm{\bar{a}}, \bm{\bar{c}}}  \max \  &
   \bm{\mu}^{\rm H} \bm{\bar{a}} \bm{\bar{a}}^{\rm H}   \bm{\mu} + \bm{\bar{c}}^{\rm H}\bm{\mu} + \bm{\mu}^{\rm H}\bm{\bar{c}} + \bm{\bar{c}}^{\rm H}\bm{\bar{c}}, \\
\text{s.t.} \quad
 & \!\!\!  \bm{b}=e^{\textsf{j}\angle(\bm{A} \bm{\mu} + \bm{c})}, \\
 & \!\!\!  \bm{\bar{a}} = \bm{b}^{\rm H}\bm{A}, \quad  \bm{\bar{c}} = \bm{b}^{\rm H}\bm{c}, \\
 & \!\!\!  |\mu_i| = 1, \ \forall i.
 \end{aligned}
\end{equation}
where the subscript $k$ is omitted for simplicity.

Note that (P4) is still non-convex and difficult to solve.
However, for any fixed $\bm{b}$, the objective takes a quadratic form,
and can be maximized by the $\bm{\mu}$ that meets the following Karush-Kuhn-Tucker (KKT) condition
\begin{equation}\label{Eq:KKT}
 \bm{\bar{a}} \bm{\bar{a}}^{\rm H}   \bm{\mu} + \bm{\bar{c}} = \text{diag}(\bm{\nu}) \bm{\mu},
\end{equation}
where $\mu_i^* \mu_i = 1$, and $\bm{\nu}=[\nu_1 \dots \nu_M]^{\rm T}$ is the Lagrangian multiplier.
Hnece, the fixed-point iteration method can be developed by
\begin{equation}\label{Eq:KKT}
 \bm{\mu}^{(q+1)} = e^{\textsf{j}\angle (\bm{\bar{a}}\bm{\bar{a}}^{\rm H}   \bm{\mu}^{(q)} + \bm{\bar{c}})},
\end{equation}
where $q$ is the iteration number.

Based on the fixed-point iteration, the objective value of (P4) is increased by fixing $\bm{b}$ and solving $\bm{\mu}$.
Reversely, an optimal $\bm{b}$ can be determined by (\ref{Eq:b}), given a fixed $\bm{\mu}$, to further maximize the objective value.
Thus, an alternating ascent method can be used to obtain a suboptimal solution to (P4) by iteratively updating the values of $\bm{\mu}$ and $\bm{b}$.
The details of the proposed method are presented in Algorithm 1.
The computational complexity is mainly contributed by matrix-vector multiplication,
which in the worst case is given by $\mathcal{O}\big(Q_1(NM+Q_2M^2)\big)$.

\begin{algorithm}[t!]
\caption{Fixed-point based alternating ascent method}
\label{AL1}
\begin{algorithmic}[1]
\Require Coefficients $\bm{A}$ and $\bm{c}$; termination thresholds $\epsilon_1$ and $\epsilon_2$; maximum numbers of iterations $Q_1$, and $Q_2$.
\State Initialize $\bm{\mu} = \bm{1}$, $\bm{b}= e^{\textsf{j}\angle(\bm{A} \bm{\mu} + \bm{c})}$, and $q_1=0$
\For{$\delta_1 \ge \epsilon_1$ and $q_1 < Q_1$}
\State $q_1 = q_1+1 $
\State $s = | \bm{\mu}^{\rm H}\bm{b}|^2$
\State Set $q_2=0$, $\delta_2 = 2\epsilon_2$, $\bm{\bar{a}} = \bm{b}^{\rm H}\bm{A}$, and $\bm{\bar{c}} = \bm{b}^{\rm H}\bm{c}$
\For{$\delta_2 \ge \epsilon_2$ and $q_2 < Q_2$}
  \State $q_2 =q_2+1 $
  \State $\bm{\eta} = \bm{\mu}$
  \State $\bm{\mu} = e^{\textsf{j}\angle (\bm{\bar{a}}\bm{\bar{a}}^{\rm H}   \bm{\mu} + \bm{\bar{c}})}$
  \State $\delta_2 = \|\bm{\eta}-\bm{\mu}\|_2^2 $
\EndFor
\State $\bm{b}= e^{\textsf{j}\angle(\bm{A} \bm{\mu} + \bm{c})}$
\State $\delta_1 = |s - | \bm{\mu}^{\rm H}\bm{b}|^2|$
\EndFor \\
\Return $\bm{\mu}$
\end{algorithmic}
\end{algorithm}

With the fixed-point based method, (P2) is solved by the following procedures: i. obtaining $\bm{\mu}_k$ based on Algorithm 1, ii. computing $\bm{b}_k$ with (\ref{Eq:b}), and iii. determining $t_k$ by (\ref{Eq:t}).

\subsection{Hardware Efficiency Gains of RRAs}
 RRAs are implemented based on tunable electromagnetic materials,
 and can provide adaptive beamforming capabilities
 without requiring additional phase shifters or expensive transmit/receive modules.
 Thus, the hardware complexity, mass, cost, and power consumption are substantially reduced \cite{RRA2,RRA3}.
 Specifically, to transmit $N_\text{s}$ independent data streams with an $N$-element array $(N \gg N_\text{s} )$,
 FDAAs require as many as $N$ RF chains for digital beamforming,
 while RRAs need only $N_\text{s}$ RF chains ($N_\text{s}=1$ in this work).
 Hence, RRAs present considerable hardware efficiency gains over FDAAs.

\section{Simulation Results}

A RIS-aided offshore communication system is investigated, where the coastal BS located at $(0,0)$,
deploys an 8$\times$8 RRA ($N=64$) with one RF feed,
and an 8 $\times$ 8 shipborne RIS ($M=64$) is placed at $(10 \text{km},0)$ in costal waters.
 The parameters for the arrays are set as $d/\lambda = 0.5$, $z^{\rm RF}=\sqrt{N}d$, and $q=6.5$.
 There are 4 vessels ($K=4$) uniformly and randomly distributed in the area
 $\mathcal{A} = \{(x,y):x^2+y^2 \leq l_\text{max}^2, x \geq 0\}$ with $l_\text{max}= 30$ km.
 The antenna heights of the BS, the RIS and the vessels are set to 50 m, 15 m, and 10 m, respectively,
 so that the vessels can be covered within the radio horizon distance given by
 $d_\text{LOS} = 4.1 (\sqrt{h_t} + \sqrt{h_r})$ \cite{Mari4}.
 The noise power spectral density is $-$170 dbm/Hz,
 the carrier frequency is 5.8 GHz, and the bandwidth for transmission is 20 MHz \cite{Mari1}.
 The involved channels are generated from (\ref{Eq:pathloss}) and (\ref{Eq:Channel}),
 with a Rician factor of $\alpha=10$ dB.

 Note that the model in (P2) can be extended to the use case of FDAAs by changing the original constraints on $\bm{R}$ and $\bm{b}_k$ into $\bm{R} = \frac{1}{\sqrt{N}} \bm{I}_N$ and $\|\bm{b}_k\|_2^2 \leq N$.
 Also, the scenario without the remote RIS is presented by setting $\bm{\mu}_k = \bm{0}$.
 Thus, the performance of the proposed design with both the RRAs and the RIS using Algorithm \ref{AL1}
 (labeled as `RRAs + RIS + AL1') can be compared with the following four benchmarks:
\begin{enumerate}
\item `FDAAs only + Optimal':
the optimal design with only the FDAAs deployed at the BS,
where $\bm{\mu}_k = \bm{0}$, and $\bm{b}_k = \sqrt{N}\frac{\bm{h}_k}{\|\bm{h}_k\|_2} $.
\item `FDAAs + RIS + SDR \cite{RIS4}':
the design with both the FDAAs and the RIS using the semidefinite relaxation (SDR) method,
where the  generic method is implemented with a large number of randomizations \cite{RIS4}.
\item `FDAAs + RIS + AL1':
the design with both the FDAAs and the RIS using Algorithm \ref{AL1},
where $\bm{\mu}_k$ is obtained by Algorithm \ref{AL1},
and $\bm{b}_k =\sqrt{N} \frac{\bm{F}^{\rm H} \bm{G}_k \bm{\mu}_k + \bm{h}_k}{\|\bm{F}^{\rm H} \bm{G}_k \bm{\mu}_k+ \bm{h}_k\|_2} $.
\item `RRAs only + Optimal': the optimal design with only the RRAs deployed at the BS,
where $\bm{\mu}_k = \bm{0}$, and $\bm{b}_k = e^{\textsf{j}\angle(\bm{R}^{\rm H}\bm{h}_k)}$.
\end{enumerate}
It should be noted that, in order to provide a fair comparison of the beamforming performances achieved by RRAs and FDAAs,
the intrinsic power consumption of the antenna arrays is not included in the simulation.
In practical systems,
the performance of RRAs with lower power consumption can be further enhanced, compared to that of FDAAs,
given the same transmission power.

To evaluate the performances under diversified rate constraints,
the rate requirements of the vessels are randomly and uniformly generated by
$\beta_k \sim \mathcal{U}(0, \log_2 (1+ \frac{\rho_\text{max}NL(l_\text{max})}{\sigma^2}) $,
instead of being constant over all $k$ as in \cite{RIS3,RIS4}.
We change the transmission power $\rho$ from $\rho_\text{min}=35$ dBm to $\rho_\text{max}=55$ dBm,
and define the service reliability (SR) as follows
\begin{equation}
\text{SR}=\frac{n_\text{SR}}{n_\text{Sim}},
\end{equation}
where $n_\text{Sim}$ is the total number of the simulation snapshots,
and $n_\text{SR}$ is the number of the snapshots wherein the rate requirements are satisfied.
The average ESR is then computed from all the $n_\text{SR}$ snapshots.

\begin{figure}[t!]
\begin{center}
\includegraphics[width=.9\linewidth]{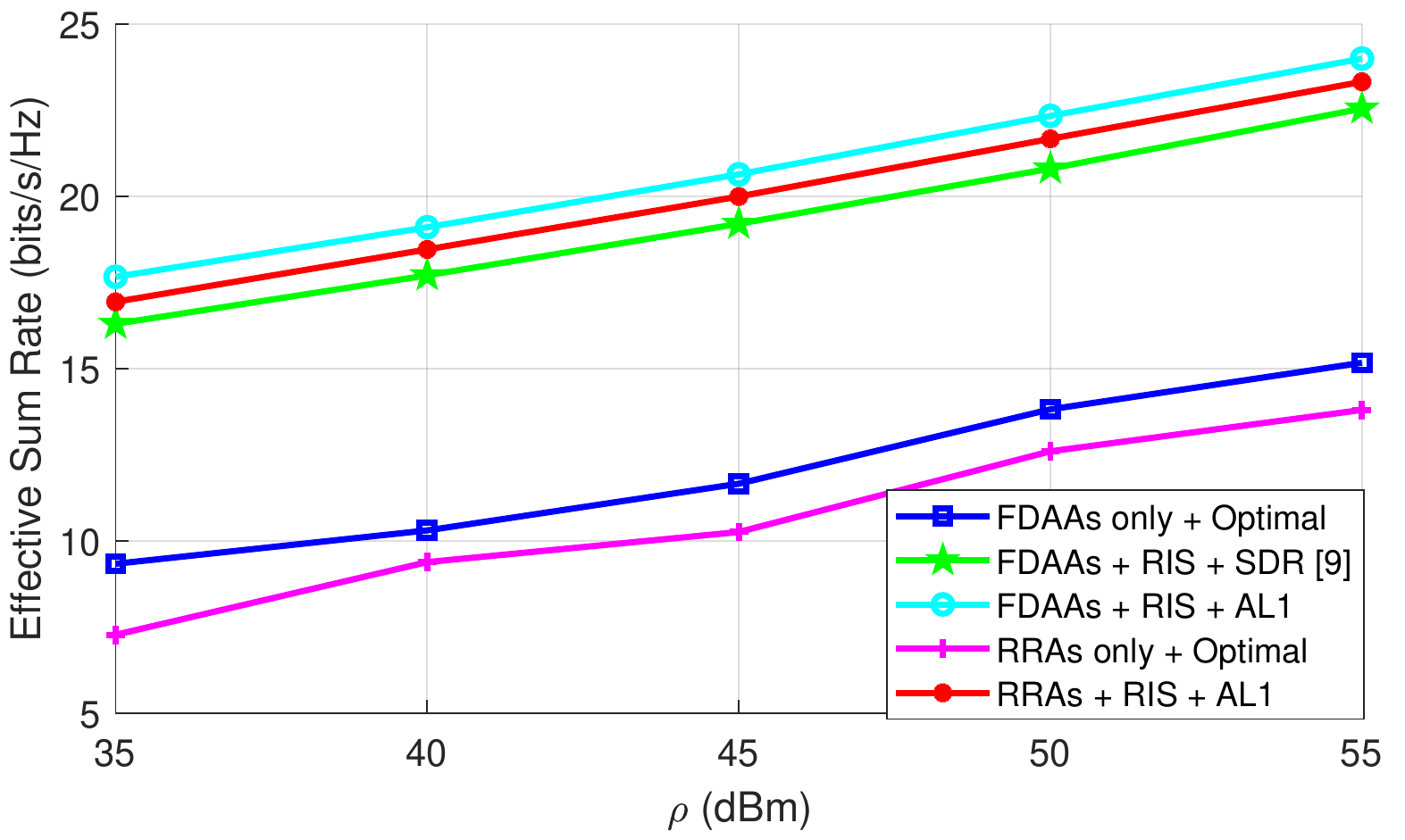}
\end{center}
\vspace*{-4mm}
\caption{ESR comparisons among various designs, where $N=8\times 8$, $M=8\times 8$, and $K=4$.}
\label{Fig3}
\end{figure}

The ESRs obtained by different schemes are compared in Fig.~\ref{Fig3}.
Observe that the ESRs are substantially enhanced by the remote RIS with either FDAAs or RRAs deployed at the BS.
The comparison between Benchmarks 2) and 3) shows that the proposed fixed-point based method is effective to obtain a suboptimal solution.
Moreover, compared to Benchmark 3), the proposed design with RRAs that reduces much implementation cost,
has only an acceptable performance loss.

\begin{figure}[t!]
\begin{center}
\includegraphics[width=.9\linewidth]{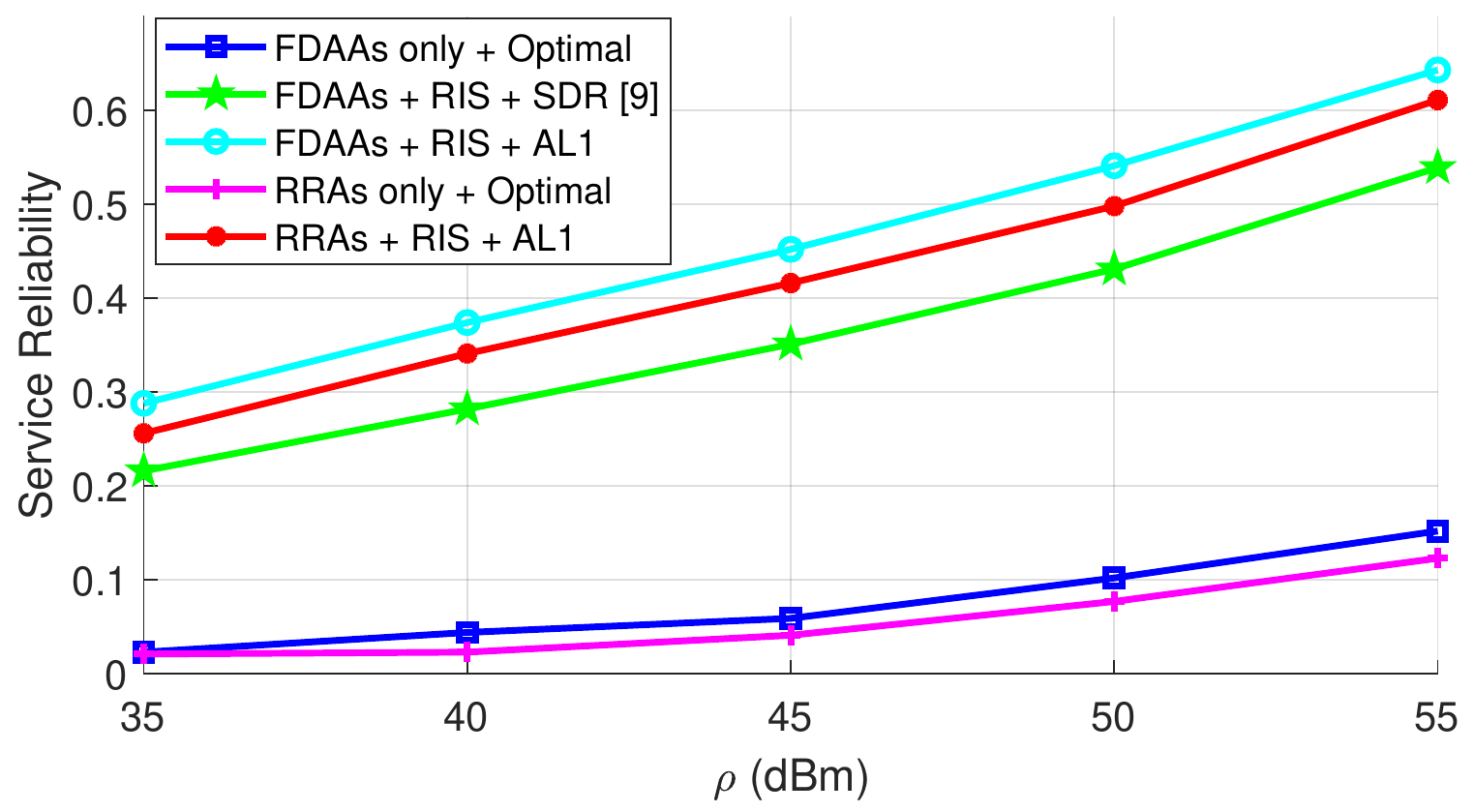}
\end{center}
\vspace*{-4mm}
\caption{SR comparisons among various designs, where $N=8\times 8$, $M=4\times 4$, and $K=4$.}
\label{Fig4}
\vspace*{-1mm}
\end{figure}

The SRs of various designs are demonstrated in Fig.~\ref{Fig4}.
Note that the SRs are improved with increased transmission power.
The gap between benchmarks 1) and 3), together with that between benchmark 4) and the proposed design using RRAs,
indicates that the RIS has considerably extended the coverage area
since the rate constraints are more likely to be satisfied, given the same power at the BS.

\begin{figure}[t!]
\begin{center}
\includegraphics[width=.9\linewidth]{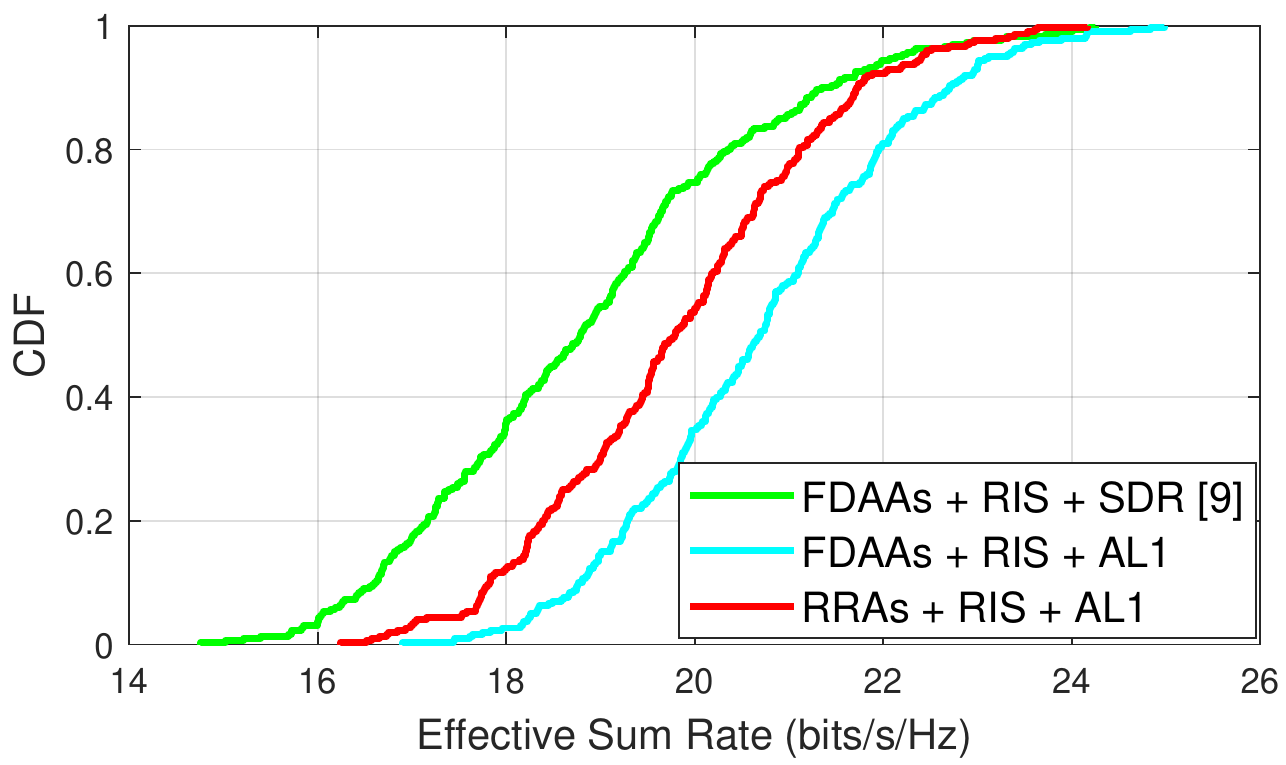}
\end{center}
\vspace*{-4mm}
\caption{CDF comparisons with $\rho=45$ dBm, where $N=8\times 8$, $M=8\times 8$, and $K=4$.}
\label{Fig5}
\vspace*{-1mm}
\end{figure}

The cumulative distribution functions (CDFs) of ESR with $\rho=45$ dBm are shown in Fig.~\ref{Fig5},
where the proposed scheme with both RRAs and RIS is compared to Benchmarks 2) and 3).
The results show that Algorithm \ref{AL1} achieves a favorable performance,
and the proposed RRA-based scheme provides a practical tradeoff between attainable ESR performance and hardware cost,
over the costly FDAA-based design.

\section{Conclusion}
The applications of the RRAs and the RIS have been investigated for the offshore system to provide a cost-effective coverage.
The mathematical model has been formulated for jointly optimizing the service time allocation and beamforming design.
The optimal allocation scheme has been derived, and an efficient fixed-point based method has been developed to obtain suboptimal beamforming solution.
Numerical results have demonstrated that
the ESR and SR performances are considerably improved with the aid of the RIS,
and the proposed scheme using the hardware-efficient RRAs
has only a slight performance loss, compared to its FDAA-based counterpart,
which shows practical utility for offshore communications.

\end{document}